\documentclass[sigconf]{acmart}

\setcopyright{none}
\renewcommand\footnotetextcopyrightpermission[1]{}
\acmConference{}{}{} % remove from arxiv
\acmBooktitle{} % remove from arxiv
\acmDOI{} % remove from arxiv
\acmISBN{} % remove from arxiv
\acmPrice{} % remove from arxiv
\usepackage{xparse}

\newcommand{\medianHaasSingleCmdSec}{0.18}

\newcommand{\nHaasLoggedIn}{255,728}
\newcommand{\nHaasProbeCatBinaryIPs}{200}
\newcommand{\nHaasProbeCatBinarySessions}{1,064}

\newcommand{\nHaasUniqueIPs}{6,579}

\newcommand{\nInteractive}{179}
\newcommand{\nOneShot}{176,256}
\newcommand{\nProbeBaseSixtyFourIPs}{4}
\newcommand{\nProbeBaseSixtyFourSessions}{2,166}
\newcommand{\nProbeCatBinaryIPs}{2}
\newcommand{\nProbeCatBinarySessions}{2}
\newcommand{\nProbeClassicIPs}{1}
\newcommand{\nProbeClassicSessions}{4}
\newcommand{\nProbeSimIPs}{9}
\newcommand{\nProbeSimSessions}{2,178}

\newcommand{\nRefused}{1,187}
\newcommand{\nSensors}{11}
\newcommand{\nSessions}{177,622}
\newcommand{\nUniqueIPs}{1,997}
\newcommand{\nUniqueOneShotCommands}{9,384}
\newcommand{\nUniqueUsers}{1,427}

\newcommand{\pctBackendPrimary}{88.5}

\newcommand{\pctHaasSingleCmdOfCmdBearing}{92.67}
\newcommand{\pctInteractive}{0.10}
\newcommand{\pctOneShot}{99.23}
\newcommand{\pctOneShotMax}{99.70}
\newcommand{\pctOneShotMin}{97.60}
\newcommand{\pctRefused}{0.67}
\newcommand{\pctTopTenOneShot}{41.59}

\newcommand{\yearHaasLongMin}{2017}
\newcommand{\yearHaasLongMax}{2026}
\newcommand{\nHaasLongSessions}{433,132,596}

\begin{document}

\title{Ghost Without Shell: Measuring Non-Interactive SSH Attacks on Honeypots}

\author{Veronica Valeros}
\affiliation{%
  \institution{Czech Technical University in Prague}
  \city{Prague}
  \country{Czechia}
}
\email{valerver@fel.cvut.cz}

\author{Muris Sladić}
\affiliation{%
  \institution{Czech Technical University in Prague}
  \city{Prague}
  \country{Czechia}
}
\email{sladimur@fel.cvut.cz}

\author{Sebastian Garcia}
\affiliation{%
  \institution{Czech Technical University in Prague}
  \city{Prague}
  \country{Czechia}
}
\email{sebastian.garcia@agents.fel.cvut.cz}

\begin{abstract}
Cyber deception research has focused on improving honeypot deception capabilities to increase attacker engagement and extend their interactions to collect more and better intelligence. For SSH honeypots, this relies on the assumption that attackers log in, open a shell, and type. We tested whether this still held by deploying eleven SSH honeypots that served both interactive and non-interactive session requests for fifteen days. We collected \nSessions{} authenticated sessions and validated our results against an independent Cowrie dataset over the same time window. We found that \pctOneShot\% of sessions were non-interactive. Interactive sessions account for only \pctInteractive\%. The same pattern held in the comparative third-party dataset used for evaluation. This finding is important because a honeypot that focuses on interactive shells or evaluates success based on session length and the number of commands can miss most authenticated attacks and draw the wrong conclusions about what attackers do after login.
\end{abstract}
\maketitle
\pagestyle{plain} % remove from arxiv

\section{Introduction}
% assumption we go after
Much of the SSH deception measurement and literature implicitly assumes that after authentication, attackers open an interactive shell to type commands~\cite{munteanuAttacksComeThose2025}. This view is common in SSH, and also Telnet honeypot work~\cite{abdouWhatLiesAnalyzing2016, srinivasaInteractionMattersComprehensive2022}, with only a few studies considering non-interactive sessions~\cite{koutskyMonitoringAnalysisCyber}. This assumption also motivates the design of recent LLM-based SSH honeypots, which focus on simulating a convincing shell dialogue~\cite{sladicLLMShellGenerative2024,otalLLMHoneypotLeveraging2024,fanHoneyLLMLargeLanguage2025}, as well as motivate the metrics used to evaluate deception, which often treat engagement time and interaction volume as proxies for success~\cite{javadpourComprehensiveSurveyCyber2024}.

However, this assumption has not been consistently checked, since SSH also supports non-interactive mode and file-transfer mode. In non-interactive mode, a client can authenticate, issue a single command via an \texttt{exec} request, read the output, and disconnect without the server ever allocating a shell. In file-transfer mode, the attacker uses scp or the sftp protocol to try to transfer files. Most SSH honeypot research reports ``post-authentication commands" but does not distinguish whether they arrived through an interactive \texttt{shell} or a non-interactive \texttt{exec} request~\cite{munteanuAttacksComeThose2025,koniarisAnalysisVisualizationSSH2013,fraunholzDataMiningLongTerm2017,zuzcakCausalAnalysisAttacks2019}. When these modes are mixed, measurements can overstate interactive behavior and understate automated non-interactive campaigns, and deception designs can optimize for dialogue that most attackers never request.

Our research studies and quantifies this phenomenon of non-interactive sessions by deploying advanced LLM-based SSH honeypots on the Internet for 15 days and analyzing which attacks use interactive and non-interactive sessions and for what. Our methodology first involves developing a variant of the AdvancedShelLM LLM-based SSH honeypot~\cite{advshellm}, making it capable of distinguishing these two session types. Second, we analyze the attacks collected during the period, extracting session commands and answers. Finally, we validate our findings against an independent Honeypot-as-a-Service (HaaS) Cowrie honeypot network operated by CZ.NIC over the same window~\cite{HoneypotServiceHome}.

Results show that \pctOneShot\% of authenticated sessions are non-interactive \texttt{exec} requests, typically completing in under a second. Interactive shell sessions only account for \pctInteractive\%, and file-transfer accounts for \pctRefused\%. This overturns the common assumption that authenticated SSH honeypot traffic is mostly interactive, and it means that shell-only honeypots and engagement-based evaluations observe a small and biased slice of post-login behavior. A non-interactive session has, by definition, only one command, which must accomplish its purpose immediately, giving the honeypot exactly one chance to respond.

Results also show that some non-interactive commands seem to be \emph{verification probes}, meaning that they test whether a target is real or a honeypot before the attacker invests further. We observe \nProbeSimSessions{} simulated-shell probes from \nProbeSimIPs{} source IPs and \nProbeClassicSessions{} classic honeypot fingerprinting sessions from \nProbeClassicIPs{} IP, while prompt-injection strings and explicit AI/LLM mentions are absent. For LLM-based SSH honeypots in particular, success depends less on how long an attacker stays connected and more on whether the honeypot survives verification.

This paper makes three contributions:
\begin{itemize}
    \item A measurement of which SSH mode attackers use after login, showing that most send a single non-interactive \texttt{exec} command rather than opening a shell, confirmed on an independent Cowrie dataset.
    \item An analysis of the nature and characteristics of these \textit{verification probes}.
    \item An argument that SSH deception research should measure whether each answer is correct and whether the honeypot passes these checks, not only how long an attacker stays.
\end{itemize}

\section{Background and Related Work}

\paragraph{Measuring SSH attacks.} The largest recent measurement of SSH attacks is by Munteanu et al., who analyzed 546 million sessions collected over three years by 221 Cowrie honeypots~\cite{munteanuAttacksComeThose2025,CowrieCowrie2026}. They classify sessions by what commands do, separating commands that alter the honeypot state from those that do not. Like most studies in this line, they do not report how commands arrive, that is, whether the attacker opened an interactive shell or sent a single exec request. That mode split is the axis this paper measures.

\paragraph{LLM-based honeypots.} Since shelLM demonstrated that an LLM can simulate a believable Linux shell~\cite{sladicLLMShellGenerative2024}, the idea has grown into a research line of its own. Examples include interactive SSH honeypots~\cite{otalLLMHoneypotLeveraging2024,fanHoneyLLMLargeLanguage2025,wangHoneyGPTBreakingTrilemma2024}, the VelLMes multi-protocol deception framework~\cite{sladicVelLMesHighInteractionAIBased2025}, web honeypots~\cite{ka0x4D31Galah2026}, industrial protocol emulation~\cite{vasilatosLLMPotDynamicallyConfigured2025}, and earlier GPT-based shell simulation~\cite{ragsdaleDesigningLowRiskHoneypots2023}. A recent SoK maps the area~\cite{bridgesSoKHoneypotsLLMs2026}, and evaluation work has begun to standardize how response quality is measured~\cite{weberDontStopBelievin2024}. Across this line, the design and the evaluation center on the interactive shell dialogue. VelLMes states this explicitly: its Internet deployment rejected non-interactive command execution by design~\cite{sladicVelLMesHighInteractionAIBased2025}. Our results show that this design choice discards about 99\% of the traffic sent by attackers.

\paragraph{Detection in both directions.} Honeypot fingerprinting is a mature topic, and frameworks exist to detect deployed honeypots at scale~\cite{srinivasaGottaCatchEm2023}. The arrival of LLMs opened a second front. Reworr and Volkov instrumented an SSH honeypot with prompt injections to catch LLM-driven attackers, finding eight candidates in over eight million attempts~\cite{reworrLLMAgentHoneypot2025}, and later work proposes challenge-based traps and proactive defenses against LLM agents~\cite{liuPuzzlePotChallenge2026,ayzenshteynCloakHoneyTrap}. All of this work points the detector at the attacker. The opposite direction, attackers probing whether the host itself is a simulation, has not been measured. That is the behavior we document in Section~\ref{sec:probes}.

\section{Methodology and Setup}

\subsection{Honeypot System}
We collected SSH attack traffic by modifying AdvancedShelLM, a high-interaction SSH honeypot that uses a large language model (LLM) to generate realistic command output and shell behavior~\cite{advshellm}. All our honeypots used the same fixed system prompt. The primary backend model was \texttt{gpt-oss-120b}, a locally deployed model that served \pctBackendPrimary\% of sessions. The OpenAI models gpt-5-nano and gpt-4.1-nano were used as a fallback provider during outages. The changes to the backend model do not affect our measurements, which concern the traffic attackers send to the honeypot rather than the responses it returns.

AdvancedShelLM accepts any username and password combination as valid. This is standard practice in SSH honeypot work and maximizes the engagement of attackers. AdvancedShelLM implements both SSH \texttt{exec requests}, in which the client submits a single command without allocating a terminal, and \texttt{shell channels}, in which the client exchanges commands through an allocated terminal. It does not support data transfer commands such as \texttt{scp} or \texttt{SFTP}. % Support for exec requests is essential because many SSH attack scripts never request an interactive shell.

\subsection{Deployment}
We deployed eleven AdvancedShelLM instances on different virtual servers from a single cloud provider in Frankfurt, Germany. All instances shared the same hardware profile (2~GB RAM, 50~GB disk, Ubuntu 24.04 LTS x64), ran the same software configuration, and listened on the standard SSH port (TCP/22). Each instance was assigned a distinct public IPv4 address from the provider's pool. These addresses are reused across tenants over time and are not freshly allocated for this study. The collection window spans fifteen days, from May 27th, 2026, to June 10th, 2026.

\subsection{Data Collected}
For each authenticated session, AdvancedShelLM logs a single JSONL record with the following features: sensor id, session id, start date time, end date time, duration, session type (Section~\ref{sec:session-types}), termination reason (e.g.: exit, disconnect, error), attack source IP address, source port, username, password, sequence of credential attempts; ordered list of attacker commands, honeypot responses. Raw packets are not captured, and no information about attackers is collected beyond what is sent to the honeypot.

%move to appendix?A session ends, and the session record is emitted, on one of six conditions. The attacker closes the channel cleanly with Ctrl+D (\texttt{eof}). The attacker types an explicit exit command such as \texttt{exit}, \texttt{quit}, or \texttt{logout} (\texttt{client\_disconnect}). The honeypot process raises an unhandled exception (\texttt{error}). The LLM backend fails irrecoverably (\texttt{backend\_failure}). The TCP connection is reset mid-session and no end event is recorded (termination reason \texttt{None}). Non-interactive exec sessions terminate as soon as the response is delivered, with a dedicated termination reason (\texttt{one\_shot\_complete}).

\subsection{Session Types}
\label{sec:session-types}
 
AdvancedShelLM identifies each authenticated session as one of the three types described above (interactive, non-interactive, or file-transfer) based on its internal knowledge of the protocol. Even though AdvancedShelLM uses a real SSH server, it is possible to identify the types because AdvancedShelLM was modified to implement a session proxy that captures the necessary information.

\paragraph{Non-interactive sessions.} The attacker issues a single non-in\-teractive exec request, e.g., \texttt{ssh user@server 'ls -a'}. The honeypot returns the output, and the SSH channel closes without a terminal ever being allocated. At the protocol level, the session consists of exactly one command--response exchange.
 
\paragraph{Interactive sessions.} The attacker requests a shell channel, usually after a \texttt{pty-req}, and exchanges zero or more commands through the allocated terminal. This is the traditional mode of SSH use.
 
\paragraph{File-transfer sessions.} The attacker attempts a file
transfer via \texttt{scp}, \texttt{rsync}, \texttt{sftp}, or the standard \texttt{sftp-server} paths. Our modification to AdvancedShelLM does not implement these functionalities and refuses these attempts. These sessions are recorded with termination reason \texttt{refused}.
%Note that only SFTP is an SSH subsystem; \texttt{scp} and \texttt{rsync} run over exec channels. The honeypot therefore labels a session as file transfer in two cases: the client requests the \texttt{sftp} subsystem, or the client issues an exec request whose first token matches a fixed list of file-transfer binaries (\texttt{scp}, \texttt{rsync}, \texttt{sftp}, and the standard \texttt{sftp-server} paths). These sessions are recorded with \texttt{mode = refused}, a label the honeypot emits only for file-transfer attempts; for exec-based attempts, the refused command is stored in a dedicated field. The first-token rule is conservative: exec requests that invoke a transfer tool behind a shell wrapper or compound command (e.g., \texttt{sh -c 'scp \ldots'}) are classified as non-interactive sessions, so our file-transfer share is a lower bound.

\subsection{Haas: The External Validation Dataset}
\label{sec:haas-mapping}
To check that our observations generalize beyond our own deployment, we compare them against an independent dataset collected with 4,737 Cowrie~\cite{CowrieCowrie2026} SSH honeypots operated by ``CZ.NIC" over the same time window~\cite{HoneypotServiceHome}. We called this dataset HaaS. An agreement between the two datasets would increase confidence that our main results are not artifacts of our instrumentation or deployment choices.

The HaaS dataset does not record the SSH session type, so we use expert heuristics to map its sessions to our three session types by only using the described commands in the logs. These are the heuristics applied, in priority order:
\begin{enumerate}
\item File-transfer sessions: These are those sessions whose first command token is one of the transfer programs: \texttt{scp}, \texttt{rsync}, \texttt{sftp}, or \texttt{sftp-server}).
\item Zero-command: the client logs in and sends no command. We discard these commandless sessions for this research. We do not know the reason for these zero-command sessions.
\item Single-command: exactly one command in the session with a median duration of approximately \medianHaasSingleCmdSec~seconds, which is too fast for a human to type. We consider this session non-interactive.
\item Multi-command: two or more commands in the session. We consider this session interactive.
\end{enumerate}

\paragraph{Heuristic bias.} This mapping is approximate: (i) one-command sessions can still be interactive (shell opened, one probe, disconnect), inflating ``non-interactive''; (ii) multiple \texttt{exec} requests can appear as multiple commands without an allocated shell, inflating ``interactive''; (iii) the first-token rule misses wrapped transfers and subsystem-level SFTP, so file-transfer is a lower bound.

\section{Results}
We group the results into specific findings from our data collection. Each subsection includes the validation results from the external dataset to facilitate the comparison to the reader.

\label{sec:results}
\subsection{Most Sessions Are Non-Interactive Commands}
\label{sec:oneshot}
Across the \nSensors{} sensors and the fifteen-day window we recorded \nSessions{} authenticated attack sessions from \nUniqueIPs{} unique source IP addresses, using \nUniqueUsers{} distinct usernames. Each session falls into exactly one of the three modes defined in Section~\ref{sec:session-types}: non-interactive sessions account for \nOneShot{} (\pctOneShot\%), interactive sessions for \nInteractive{} (\pctInteractive\%), and file-transfer sessions for \nRefused{} (\pctRefused\%). %sum checks up

\begin{table}[t]
\centering
\small
\caption{Session-mode breakdown across all honeypots and the full 15-day window.}
\label{tab:session-modes}
\begin{tabular}{lrr}
\hline
\textbf{Mode} & \textbf{Sessions} & \textbf{Share} \\
\hline
Non-interactive \texttt{exec} & \nOneShot{} & \pctOneShot\% \\
Interactive \texttt{shell} & \nInteractive{} & \pctInteractive\% \\
File-transfer attempts (refused) & \nRefused{} & \pctRefused\% \\
\hline
Total & \nSessions{} & 100\% \\
\hline
\end{tabular}
\end{table}
Table~\ref{tab:session-modes} summarizes the aggregate breakdown of session modes across our entire deployment window. 

\begin{figure}[t]
\centering
\includegraphics[width=\columnwidth]{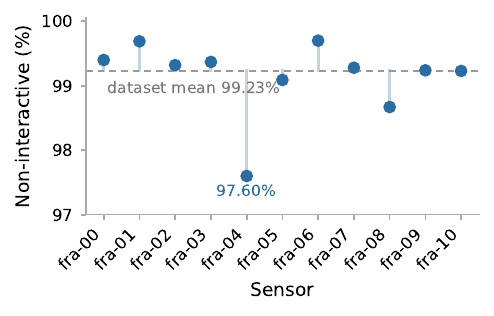}
\caption{Fraction of authenticated sessions that are non-interactive per sensor. Every sensor sits at or above \pctOneShotMin\%, with the dashed line marking the dataset mean. Fra-00 through Fra-10 are the names of the cloud computers with the honeypots.}
\Description{Bar chart with one bar per honeypot sensor, Fra-00 through Fra-10. All bars are between 97.6 and 99.7 percent, and a dashed horizontal line marks the fleet mean near 99 percent.}
\label{fig:oneshot-pct}
\end{figure}

This pattern is consistent across sensors and is not driven by a single high-volume host. Figure~\ref{fig:oneshot-pct} reports, for each of the \nSensors{} sensors, the fraction of authenticated sessions that are non-interactive. The values range from \pctOneShotMin\% to \pctOneShotMax\%.

We cross-checked this result against the CZ.NIC HaaS Cowrie dataset described in Section~\ref{sec:haas-mapping}. In the matching window it contained \nHaasLoggedIn{} logged-in sessions from \nHaasUniqueIPs{} unique source IPs. Among sessions that contained at least one command, \pctHaasSingleCmdOfCmdBearing\% contained exactly one command. This matched our observations and indicated that non-interactive activity was not specific to our deployment.

\subsection{Reconnaissance as Non-Interactive Commands}
\label{sec:oneshot-what}
The distribution of non-interactive commands is highly concentrated: the ten most frequent commands account for \pctTopTenOneShot\% of all non-interactive sessions (Table~\ref{tab:top-cmds}). Across the full dataset we observed \nUniqueOneShotCommands{} distinct non-interactive command strings, but the most common commands are largely used for reconnaissance, because they collect system and environment information without changing state. They query basic host properties (\texttt{uname}, \texttt{nproc}, \texttt{lspci}), identity (\texttt{whoami}), and uptime (\texttt{uptime}).

\begin{table}[t]
\centering
\caption{Top ten non-interactive commands by session count. The top ten cover \pctTopTenOneShot\% of non-interactive traffic.}
\label{tab:top-cmds}
% Auto-generated by analysis/analysis.py. Do not edit by hand.
\begin{tabular}{p{0.65\columnwidth}rr}
\toprule
Command & Sessions & \% \\
\midrule
\texttt{uname -s -v -n -r -m} & 13,262 & 7.52 \\
\texttt{uname -a} & 11,817 & 6.70 \\
\texttt{uname -m | awk '\{printf \$1\}'} & 9,602 & 5.45 \\
\texttt{echo ''} & 6,492 & 3.68 \\
\texttt{whoami} & 6,260 & 3.55 \\
\texttt{nproc} & 6,053 & 3.43 \\
\texttt{lspci | egrep VGA \&\& lspci | grep 3D} & 5,821 & 3.30 \\
\texttt{uname -m} & 5,804 & 3.29 \\
\texttt{export HOME=/dev/null \ldots} & 4,570 & 2.59 \\
\texttt{uptime} & 3,619 & 2.05 \\
\midrule
Top ten commands & 73,300 & 41.59 \\
All other commands (9,374 distinct) & 102,927 & 58.40 \\
No command recorded & 29 & 0.02 \\
\midrule
Total (9,384 distinct) & 176,256 & 100.00 \\
\bottomrule
\end{tabular}

\end{table}

We cross-checked command content in the external dataset, and the reconnaissance commands that are most common in our non-interactive traffic also appear in HaaS. Commands such as \texttt{uname -a}, \texttt{whoami}, \texttt{pwd}, \texttt{uptime}, and \texttt{cat /proc/cpuinfo | grep 'model name' | head -1} were common in both datasets, suggesting that the same automated campaigns reached both deployments.

\subsection{Verifying if the Target is a Honeypot}
\label{sec:probes}
Some non-interactive commands are not reconnaissance. Instead, they test whether a target behaves like a real host before the attacker invests further effort. We observe two kinds of such verification probes: checks for a simulated shell, and checks for known honeypots. Each is described below.

\subsubsection{Checks for a simulated shell.}
The first kind tests whether the remote side executes commands and returns the correct output, rather than generating plausible text. This matters for LLM-based honeypots, which can answer in a way that looks reasonable but is not consistent with actual execution.

We observed \nProbeSimSessions{} simulated-shell check sessions in total. The largest campaign used a base64 decoding command. This campaign had \nProbeBaseSixtyFourSessions{} sessions from \nProbeBaseSixtyFourIPs{} IPs and it used commands like \texttt{echo 'dGVzdA==' | base64 -d}, which should return the string \texttt{test} on a real system. Other commands included exact arithmetic (e.g., \texttt{echo \$((7*6))}), binary dumps (e.g., \texttt{cat /bin/echo}), and state/timing checks such as writing a file and reading it back or \texttt{sleep 0.1 \&\& echo done}.

We found the same class of execution-check probes in the HaaS dataset, including campaigns that issued arithmetic and system-identification sequences (e.g., \texttt{echo \$((7*6)); uname -m; cat /proc/uptime; head -1 /proc/version}). One command in this family, \texttt{cat /bin/echo}, was far more common against Cowrie (\nHaasProbeCatBinarySessions{} sessions from \nHaasProbeCatBinaryIPs{} IPs) than against our dataset (\nProbeCatBinarySessions{} sessions from \nProbeCatBinaryIPs{} IPs).

\subsubsection{Checks for a known honeypot.}
We observed \nProbeClassicSessions{} classic honeypot-fingerprinting sessions from \nProbeClassicIPs{} source IPs. These probes look for tells of Cowrie and similar systems, such as listing processes for \texttt{cowrie} or \texttt{kippo}, testing whether \texttt{/etc/passwd} is writable, and checking for known fake users.

We also checked the HaaS dataset for these fingerprinting commands. Using the same string-based detectors, we did not find corroborating classic fingerprinting sessions in HaaS. These results are discussed in Section~\ref{sec:discussion}. 

\subsubsection{What we did not find.}
We screened all \nSessions{} sessions for prompt-injection strings, explicit mentions of AI, LLMs, or named models, and fixed-string hash probes. None appeared. We repeated the same screen on the HaaS dataset and found no prompt-injection strings, no explicit mentions of AI/LLMs or model names, and no fixed-string hash probes.

\section{Discussion}
\label{sec:discussion}

Our results suggest that post-authentication SSH attack activity is better described as a stream of atomic automated checks and short reconnaissance commands than as interactive operator sessions. In our deployment, \pctOneShot\% of authenticated sessions consisted of a single non-interactive \texttt{exec} request, while only \pctInteractive\% opened an interactive shell. The CZ.NIC HaaS dataset from 4,737 honeypots shows the same pattern in the matching window. 

To see when this shift occurred, we examined the HaaS archive over multiple years. Across \nHaasLongSessions{} sessions from \yearHaasLongMin{} to \yearHaasLongMax{}, non-interactive sessions have been common since at least 2018. This suggests many attackers now use SSH in a more automated way: log in, run one quick command, and move on as part of a larger scan or exploit run.

The changes happen from month to month. The biggest changes are a +17.3 percentage-point jump in non-interactive sessions in October 2024 (from 80.1\% to 97.4\%) and a $-$13.8 percentage-point drop in interactive sessions in October 2019 (from 22.8\% to 9.0\%), computed over sessions with at least one command, as in Section~\ref{sec:haas-mapping}. In both cases, the change comes with a spike in total traffic, suggesting that a few large automated campaigns can move the monthly numbers, even though non-interactive sessions remain the majority overall.

The biggest implication of this measurement is methodological: many SSH honeypot studies implicitly treat ``post-authentication commands'' as interactive shell activity. If most authenticated traffic is non-interactive, analyses that focus on interactive or multi-command sessions examine only a small subset of attacker behavior. Likewise, a honeypot that does not implement non-interactive requests produces a dataset that reflects what the honeypot allows, not what attackers attempt. To make results comparable, SSH honeypot studies should report which SSH modes are supported (\texttt{shell}, \texttt{exec}, subsystems such as SFTP) and the share of authenticated sessions observed in each mode.

In deception design, non-interactive activities reduce the opportunity for narrative or dialogue-based techniques and shift the problem toward correctness under automation. It also challenges common evaluation metrics such as engagement time and interaction volume, because non-interactive sessions end quickly and a single TCP ``session'' may not correspond to a single actor’s activity. Recovering the intelligence value of one traditional interactive session, therefore, requires grouping many non-interactive sessions into meta-sessions per actor or campaign.

Section~\ref{sec:probes} shows that some non-interactive commands are for verification: they check whether a target executes commands correctly, maintains state, and behaves like a real shell. This is a particular challenge for LLM-based shells, which can produce output that looks plausible but is not consistent with real execution. For these commands, engagement-based metrics are not informative. A more direct evaluation is (i) whether responses match a real system for common commands, (ii) whether state changes persist when re-checked, and (iii) how often the honeypot passes probe patterns observed in the wild. 

Across both datasets, we did not observe prompt-injection strings or explicit mentions of AI or model names. The verification activity we did observe was operational: attackers tested outputs and state rather than trying to steer a model through language. This does not rule out LLM-driven attackers, but it suggests that, in these datasets, the main pressure on LLM-based deception comes from execution-checking probes rather than conversational manipulation.

\section{Limitations}
Our measurement reflects what honeypots on commodity cloud address space can see: indiscriminate, Internet-wide automated activity against hosts that accept any credentials. Attackers targeting a specific production system may still behave interactively, and our results do not cover them. The honeypot also shapes its own data: it refuses file transfers, and a campaign that probes a host before engaging further may stop after a failed check, which would inflate the non-interactive share by the very mechanism we document.

\section{Future Work}
Our future work includes three directions. First, a longer deployment across more providers and address space, with file-transfer support, to test how much the vantage point and the honeypot interface shape the traffic.
Second, a controlled experiment in which honeypots deliberately pass or fail verification probes, to measure whether surviving a check changes what attackers send next. Third, grouping non-interactive sessions into meta-sessions per campaign and developing the probe-passing metric into a benchmark against which LLM-based honeypots can be evaluated before deployment.

\section{Conclusion}
Our measurements show that most authenticated SSH attacks do not use an interactive shell. In our deployment, \pctOneShot\% of sessions consisted of a single non-interactive \texttt{exec} command and completed at machine speed; the independent Cowrie dataset showed the same pattern. This implies that shell-centric honeypot deployments and analyses systematically miss the dominant post-login behavior. Within this non-interactive traffic, we observed verification probes, including probes that test whether a target actually executes commands. These findings suggest two practical consequences for SSH deception: honeypots should support non-interactive \texttt{exec} requests to capture most authenticated traffic, and evaluation should emphasize per-command correctness and robustness to automated verification rather than session length or dialogue turns.

% acmart's \begin{acks}...\end{acks} environment hides itself automatically
\begin{acks}
This work was supported by the Ministry of Education, Youth and Sports of the Czech Republic through the e-INFRA CZ (ID: 90254).
\end{acks}

\bibliographystyle{ACM-Reference-Format}
\bibliography{references}

\appendix

\section{Ethical Considerations}
Our honeypots logged only the traffic that attackers sent to them. We collected no personal data beyond source IP addresses, which we use only in aggregate and do not publish. No human subjects were involved in this study.

\section{Generative AI Usage}
The authors used LLM assistance (Anthropic Claude Opus 4.7) to improve the language, grammar, and flow of this paper. Additionally, LLM assistance was used to code-review data-processing scripts to automate the analysis. All ideas, structure, and text were created by the authors. The authors reviewed and edited all outputs. The authors take full responsibility for the content of the paper.

\end{document}